\documentclass[sigconf, screen, nonacm, timestamp=false]{acmart}

\usepackage{tikz}
\usepackage{graphicx}

\AtBeginDocument{%
  \providecommand\BibTeX{{%
    \normalfont B\kern-0.5em{\scshape i\kern-0.25em b}\kern-0.8em\TeX}}}

\copyrightyear{2023}
\acmYear{2023}
\setcopyright{acmlicensed}
\acmConference[EASE '23]{Proceedings of the International Conference on Evaluation and Assessment in Software Engineering}{June 14--16, 2023}{Oulu, Finland}
\acmBooktitle{Proceedings of the International Conference on Evaluation and Assessment in Software Engineering (EASE '23), June 14--16, 2023, Oulu, Finland}
\acmPrice{15.00}
\acmDOI{10.1145/3593434.3593474}
\acmISBN{979-8-4007-0044-6/23/06}

\begin{document}

\title{Analyzing Maintenance Activities of Software Libraries}

\author{Alexandros Tsakpinis}
\affiliation{%
  \institution{fortiss - Research Institute of the Free State of Bavaria}
  \city{Munich}
  \country{Germany}}
\email{tsakpinis@fortiss.org}

\begin{abstract}
    Industrial applications heavily integrate open-source software libraries nowadays. Beyond the benefits that libraries bring, they can also impose a real threat in case a library is affected by a vulnerability but its community is not active in creating a fixing release. 
    Therefore, I want to introduce an automatic monitoring approach for industrial applications to identify open-source dependencies that show negative signs regarding their current or future maintenance activities. 
    Since most research in this field is limited due to lack of features, labels, and transitive links, and thus is not applicable in industry, my approach aims to close this gap by capturing the impact of direct and transitive dependencies in terms of their maintenance activities.
    Automatically monitoring the maintenance activities of dependencies reduces the manual effort of application maintainers and supports application security by continuously having well-maintained dependencies.
\end{abstract}


\maketitle

\section{Introduction}

\textbf{Situation:} Over the years, the usage of Open Source Software (OSS) has become a standard along the product life cycle due to commercial, engineering and quality reasons \cite{ebert2008open}. In fact, around 80\% of the code of an average commercial product consists of OSS components underlining the consolidated practice in the software industry \cite{pittenger2016open}. One element of OSS are libraries, which play a key role in modern software development. The idea of software libraries is that instead of writing code for a specific functionality from scratch, one would rather reuse well-tested and established code from a third-party library providing the desired functionality \cite{bauer2012structured}. Once a library is included in a project, it can be defined as a dependency \cite{cox2019surviving}. OSS dependencies may introduce vulnerabilities (e.g., Log4j\footnote{\url{https://logging.apache.org/log4j/2.x/}}) which is among the most pressing and important problems in the software industry \cite{decan2018impact}. If a vulnerability is detected, a fixed version is usually released soon after by the library's community \cite{rahkema2022swiftdependencychecker}.

\textbf{Problem:} As applications evolve independently from their dependencies, the application maintainers need to monitor the evolution of the dependencies carefully \cite{kula2014visualizing}. For instance, it can happen, that the support of a library is discontinued or suspended, such that no extensions or fixes for critical bugs can be expected anymore \cite{bauer2012structured, raemaekers2011exploring}. If a system depends directly on a library which is poorly maintained, any vulnerability in that library might be harmful for the system itself \cite{reifer2003eight}.    
In addition, depending on a library with a vulnerability affects not only all dependent projects, but also all projects and libraries that are connected via transitive links further increasing the number of affected projects \cite{pashchenko2018vulnerable}. 

In case such a problematic scenario occurs where an application depends on a library affected by a vulnerability for which there is no active community, there are two cost-intensive ways to fix this security threat. First, the application maintainers or a contracted software service provider take care of the problem by contributing to the library, i.e., develop the code for its new release \cite{pashchenko2018vulnerable, raemaekers2011exploring}. Second, the application maintainers replace the library with a different one that provides the same desired functionality \cite{bauer2012understanding, raemaekers2011exploring}. 

Another problem is, that once a library is installed, monitoring the community support and maintenance activities is currently mostly done manually \cite{cleare2018gemchecker}. With a small number of dependencies, this might not be a problem. Yet, for real world systems, various studies report a high number of direct and transitive dependencies leading to an infeasible effort to monitor their maintenance activities \cite{bauer2012structured, huang2022characterizing, kikas2017structure}.
This large number of direct and transitive dependencies, combined with the need to monitor the maintenance activities, requires a reduction in the cost of automatically monitoring the maintenance activities of dependencies in a system \cite{cox2019surviving, pashchenko2018vulnerable}.

However, most research is limited due to lack of features, labels, and transitive links and is thus not applicable in the industry \cite{pashchenko2020vuln4real}. Either, the maintenance activities of transitive dependencies are not considered at all \cite{coelho2020github}, or only the first level is included in the analysis \cite{li2022ossara}. To the best of my knowledge there is no research that fully captures the impact of direct and transitive dependencies in terms of their maintenance activities. 
Besides that, existing approaches classify libraries only based on their current status and miss to make predictions about the future \cite{coelho2020github,li2022exploring}. Predicting a library's future maintenance activity allows application maintainers to consider in advance what actions should be taken for a library showing negative signs of future maintenance activity before its maintenance support is actually discontinued. Also, they lack an evaluation that focuses on effort awareness, which is important to be applicable in industry. In my context, effort awareness measures to which degree the effort of application maintainers can be reduced when monitoring the maintenance activities of OSS libraries.

\textbf{Solution:} The aim of this dissertation is to research novel techniques that allow to automatically monitor the maintenance activities of dependencies in real world software systems. To solve this problem, I propose to divide the problem into two subtasks, namely classification and prediction. For the classification, a library should be analyzed based on its project-level features (e.g., number of commits, contributors, stars, etc.) and assigned to a class based on how active the community support is at the moment. For the prediction, relevant project-level features should be predicted for different time periods which can then be used to assign a label describing how active the community support will be in the future. For both subtasks, it is essential to consider the maintenance activities of direct as well as transitive dependencies. As a result, problematic dependencies are identified showing signs of negative maintenance activities either currently or in the future.

\textbf{Contribution:} As scientific contribution, I want to create a dataset with extended and combined maintenance activity features and labels. The proposed labels identified in the literature include \textit{active}, \textit{feature complete}, \textit{dormant} and \textit{inactive} which describe a level of maintenance activities for a library. The dataset is used to build a model that classifies libraries in terms of their maintenance activities and makes predictions about their maintenance activities for different time periods, taking into account both direct and transitive dependencies. The proposed methods shall be evaluated in a case study focusing on effort awareness.

As practical contribution, I want to reduce the effort of application maintainers to monitor the dependencies of an industrial application in terms of their maintenance activities by providing an assisting tool. Such a tool supports application security, by continuously assuring well-maintained dependencies.
\section{Motivation of use case}

\textbf{Goal:} My use case aims to automatically monitor the maintenance activities of direct and transitive OSS dependencies in industrial applications to identify OSS libraries showing suspicious maintenance activities (e.g., feature complete, dormant, inactive). 

\textbf{Actors:} The main actors are security managers and external software auditors who are responsible for ensuring that only secure OSS is used within an organization. If there is no separate security manager in an organization, my work can also be applied to assist developers directly. Note that bigger organization will benefit more from the continuously generated maintenance activity report since they have more employees to make use of the information.

\textbf{Relevance for actors:} My use case is of particular importance since most companies currently do not have a sufficient monitoring mechanism for the maintenance activities of OSS dependencies. Either the maintenance activities are monitored manually or they are not monitored at all, which is either very costly or insecure. However, the lack of such a monitoring mechanism and thereby unknown unmaintained dependencies is a potential security threat.

\textbf{Pre-conditions:} The OSS libraries used in the system under investigation should be accessible via source code or via its build definition (e.g., pom.xml). My use case is restricted to OSS libraries that are listed in publicly accessible software repositories (e.g., GitHub) to be able to collect the data for the required features.

\textbf{Triggers:} The main trigger for my use case is a new CI run. My approach would be integrated in a CI step as CLI tool which creates for every CI run a report about the current and future maintenance activity status for each OSS dependency. Alternatively, the report generation can manually be triggered by providing a single library as string or a file containing multiple libraries to the CLI tool.

\textbf{Temporal applicability:} My use case is applicable at any point within the application life cycle. For prototypes and early-stage applications without a CI pipeline the CLI tool can be used as manual trigger to evaluate if a selected library has sufficient maintenance activities before being installed. For all other applications that use a CI pipeline, my tool can be integrated as a separate CI step, regardless of which phase of the application life cycle they are in.

\textbf{Basic flow:} The basic flow consists of two steps. First, each library shall be classified based on the current maintenance activities. Second, the maintenance activities of all libraries are predicted for various time periods in the upcoming months. Hence, current or future suspicious OSS dependencies can be identified. 

It is important that the classification and prediction tasks are not only applied on direct dependencies but also take transitive dependencies into account. This additional information about transitive dependencies might not be of interest to libraries that are already classified as suspicious due to their own maintenance activities, since the aggregated result of the transitive dependencies would not change the overall result anyway. However, imagine the scenario where the maintenance activities of a library seem fine, however there are transitive dependencies which would be identified as suspicious. Such a library should be marked as suspicious and be analyzed more in detail by the responsible person since a vulnerability in a transitive dependency can be as dangerous as in the root library. This means that a library is marked as unsuspicious only if all its transitive dependencies also have no negative signs.

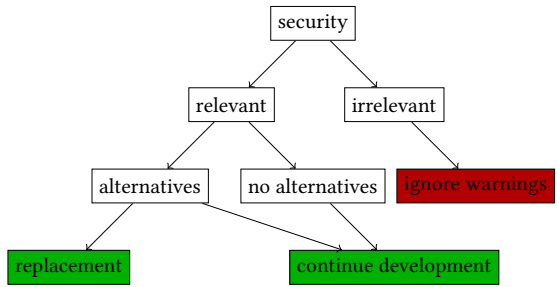
\begin{figure}[ht]
    \centering
        \scalebox{.9}{
            \begin{tikzpicture}[node distance=17mm, main/.style = {draw, minimum height=0.5cm}] 
                \node[main] (1) {security};
                \node[main] (2) [below left of=1] {relevant};
                \node[main] (3) [below right of=1] {irrelevant};
                \node[main] (4) [below left of=2] {alternatives};
                \node[main] (5) [below right of=2] {no alternatives};
                \node[main, fill=black!30!red] (6) [below right of=3] {ignore warnings};
                \node[main, fill=black!30!green] (7) [below left of=4] {replacement};
                \node[main, fill=black!30!green] (8) [below right of=5] {continue development};
        
                \draw[->] (1) -- (2);
                \draw[->] (1) -- (3);
        
                \draw[->] (2) -- (4);
                \draw[->] (2) -- (5);
        
                \draw[->] (3) -- (6);
        
                \draw[->] (4) -- (7);
                \draw[->] (4) -- (8);
        
                \draw[->] (5) -- (8);
        
            \end{tikzpicture}}
    \caption{Possible scenarios for library reported as suspicious}
    \label{scenarios-for-reported-library}
\end{figure}

\textbf{Post-conditions:} After suspicious dependencies have been reported as part of the basic flow, it remains to describe what the responsible person might do with this information. If a library is reported as suspicious, three actions can typically be taken, namely ignore warnings, replacement of the library if alternatives exist, or continuation of the development. Continuing the development can either take place in a private forked project or the code can be contributed to the OSS project. Depending on an organization's security requirements for its application(s), different scenarios are possible if a library is reported as suspicious, as shown in Figure \ref{scenarios-for-reported-library}.

The maintenance activity report can be seen as a critical factor in assuring application security by initiating one of the security-enhancing actions marked in green. Which action to be taken depends on the specific context and must be decided by the responsible person. The security manager could use the reported information to plan library replacement/development tasks, or to conduct a risk assessment of how integrated the reported libraries are into an application if no resources are available to address the security threat. An external software auditor could use the reported information about current or future suspicious libraries as part of its assessment.

\textbf{Expected Benefits:} There are two main benefits of automatically monitoring the maintenance activities of OSS libraries in a system. First, it reduces the effort of application maintainers since they would only need to analyze a small portion of the whole library set which increases exponentially when considering direct and transitive dependencies.
Second, after suspicious libraries are identified and countermeasures have been initiated accordingly, it is possible to continuously have well-maintained libraries in a system preventing the situation of a library being affected by a vulnerability without active community support.
\section{Research Questions}

The following four research questions (RQ) result from the situation and problems mentioned above:

\textbf{RQ1:} Which features and labels characterize OSS libraries in terms of their maintenance activities?
 
\textbf{RQ2:} How performant is the classification of OSS libraries regarding their maintenance activities considering direct and transitive dependencies in terms of precision and recall?

\textbf{RQ3:} How performant is the prediction of maintenance activities for OSS libraries for different time periods considering direct and transitive dependencies in terms of precision and recall?

\textbf{RQ4:} To what extent can the effort of application maintainers spent to monitor the OSS libraries of an industrial application in terms of their maintenance activities be reduced using a tool?
\section{Related Work}
    Below, the most relevant literature is summarized for each RQ.
    
\subsection{RQ1: Characterizing Maintenance Activities}

    Two recent studies show that community support is among the top three \cite{de2018empirical} or most important \cite{li2022exploring} factors for developers when selecting and adopting third-party libraries and OSS. Considering these factors is based on the assumption that the maintenance activities, which already play an important role during the selection and adoption of the library, must also be continuously monitored after the integration. Due to space limitations, concrete features for maintenance activities are not mentioned here, as a first literature review found over 30 sources. As maintenance activity labels, the states active, feature complete, dormant and
    inactive have been identified \cite{coelho2020github, li2022ossara, decan2017empirical, khondhu2013all, valiev2018ecosystem, cox2019surviving}. To better understand the underlying data, \citeauthor{choi2022attack} proposed an unsupervised clustering approach ($k$-means) \cite{choi2022attack}. Therefore, $k$ could be set to four according to the identified maintenance activity labels. The main gap of RQ1 is an empirical evaluation about existing maintenance activity features and labels to better understand how to characterize OSS libraries.

\subsection{RQ2: Classifying Maintenance Activities}

    To identify unmaintained OSS libraries, \citeauthor{coelho2020github} applied a random forest algorithm using 13 project-level GitHub features \cite{coelho2020github}. However, the classification is only based on project-level features and does not take direct and transitive dependencies into account, which is a significant gap to be applicable in an industrial context. To address this gap, the aggregated abandonment risk for every OSS library in a system was computed by calculating the risk for each direct dependency including their direct dependencies. That means, transitive dependencies for the first level were included, however all transitive links are still missing \cite{li2022ossara}. One way to integrate the effect of transitive dependencies is by using different impact measurements, such as degree centrality, in-degree centrality, out-degree centrality and eigenvector centrality \cite{banks2022measuring}. Similarly, the page rank algorithm was applied to identify packages in decline based on the centrality of packages \cite{mujahid2021toward}. This method could be adjusted for my research by using the maintenance activity of each library as measure for the page rank algorithm. A similar problem with transitive OSS dependencies is known from license violations which can be automatically detected \cite{duan2017identifying} and addressed by implementing best-practices for open-source governance and compliance in companies \cite{harutyunyan2019getting}. Interestingly, there is an explicit call for future work to build a precise model for the automatic identification of unmaintained libraries \cite{pashchenko2018vulnerable, pashchenko2020vuln4real} which motivates my work, but neglects to distinguish between the different maintenance activity states identified \cite{coelho2020github, li2022ossara, decan2017empirical, khondhu2013all, valiev2018ecosystem, cox2019surviving}. I plan to address this gap by moving from a binary to an multi-class approach.
    
\subsection{RQ3: Predicting Maintenance Activities}

    Different studies predicted specific features for different topics, such as health related features \cite{xia2022predicting, li2022ossara} or popularity measures \cite{bidoki2018cross, sahin2019predicting}. Predictions for multivariate maintenance activity features are missing so far. Statistical algorithms such as logistic regression, k-nearest neighbors, support vector regression, linear regression and regression trees \cite{li2022ossara, xia2022predicting}, but also neural network based algorithms, such as LSTM RNNs \cite{bidoki2018cross, sahin2019predicting} were applied. The prediction periods range from 1 to 30 days \cite{li2022ossara, xia2022predicting, sahin2019predicting, bidoki2018cross}, 1 to 6 months \cite{li2022ossara, xia2022predicting, bidoki2018cross} and up to 12 months and longer \cite{xia2022predicting, bidoki2018cross}. The main gap of RQ3 is the integration of transitive links into the prediction approach since current approaches are based on project-level features only.

\subsection{RQ4: Case Study on Effort Awareness}

    Available case studies with industry partners evaluated the feasibility and performance to identify suspicious OSS libraries \cite{li2022ossara, pashchenko2018vulnerable, pashchenko2020vuln4real}, but have not examined the extent to which effort can be reduced when comparing a manual and tool-based monitoring approach. In fact, there is a lack of publicly available tools for evaluating OSS libraries applicable for industrial environments \cite{lenarduzzi2020open}. Even though there are multiple tools\footnote{\url{https://debricked.com/}}\footnote{\url{https://tidelift.com/}}\footnote{\url{https://snyk.io/}}\footnote{\url{https://endorlabs.com/}} managing OSS libraries in terms of SBOM generation, vulnerability identification or security reporting, the aspect about maintenance activities is either limited, too simplistic or not transparent. To visualize the effect of direct and transitive dependencies, \citeauthor{chen2020rem} built a tool which highlights problematic dependencies induced via transitive links within a dependency network using publicly available NPM features \cite{chen2020rem}. For continuous monitoring, my approach could be integrated into an automated building process, similar to \textit{AuditJS}\footnote{\url{https://www.npmjs.com/package/auditjs}} or \textit{Dependabot}\footnote{\url{https://github.com/dependabot}}.
\section{Research Plan}
Below, the contribution and methods are summarized for each RQ.

\subsection{RQ1: Characterizing Maintenance Activities}

I plan to extend the state-of-the-art features and labels characterizing OSS libraries in terms of their maintenance activities by following a mixed-methods approach that combines a quantitative and qualitative study which increases the validity of the results and provides a deeper, broader understanding of the topic under research \cite{mckim2017value}. 

As quantitative study, a literature review will be performed where state-of-the-art literature is analyzed looking for features to characterize OSS libraries in terms of their maintenance activities (e.g., number of commits, number of (core) contributors, average response time to issue). To extend the state-of-the-art features, additional features are extracted from the limitations or future work sections of corresponding papers that are expected to have a promising impact on characterizing OSS projects in terms of their maintenance activities, but have not yet been evaluated.
To the best of my knowledge, no work has tried whether the combination of existing features could improve the performance of the classification. Besides that, multiple studies also indicate that additional features could further improve the performance of their approaches \cite{coelho2020github,li2022ossara}.

In addition, I plan to extend the labels of existing approaches during the quantitative study since to the best of my knowledge only datasets with binary labels, such as maintained / active vs. unmaintained / inactive / abandoned / archived, exist at the moment \cite{coelho2020github, li2022ossara, decan2017empirical, khondhu2013all}. However, there are two more states mentioned in the literature which should be handled separately, namely dormant and feature complete. A dormant project has temporarily suspended its maintenance activities but is not yet defined as unmaintained \cite{valiev2018ecosystem}. A feature complete project will most likely show signs of an unmaintained or dormant project, but does not require active maintenance since the project is complete it terms of its functionality and may never need to be modified again (e.g., NPM’s escape-string-regexp\footnote{\url{https://www.npmjs.com/package/escape-string-regexp}}) \cite{cox2019surviving}. Even though a feature complete project shows development activities like an unmaintained or dormant project, such a project usually has a community in the background which would act in case of an emergency (e.g., detected vulnerability).

It is necessary to handle projects in these two additional states differently compared to maintained or unmaintained ones. A feature complete dependency could safely be kept, even though it shows signs of being unmaintained. With existing binary classification approaches, feature complete projects would be falsely counted as unmaintained, since they often have no commit activity for over a year. Such a threshold, based on the last commit activity, is commonly used to identify unmaintained projects leading to false positives \cite{khondhu2013all}. In contrast, previous approaches would falsely classify dormant projects as maintained because they often consider projects as unmaintained that are either archived or have it explicitly defined in their README, leading to false negatives \cite{coelho2020github}.

To be able to label each repository in terms of its maintenance activities, the literature should be examined for definitions of labels such as active, feature complete, dormant and inactive. Besides the four labels already identified in the literature, more labels might be found during the literature review of the quantitative study. Since most label definitions are based on simplistic and limited assumptions, an approach should be elaborated that provides a realistic labeling strategy while resolving the limitations.

After the quantitative study, a qualitative study shall be conducted using two target groups to generate new ideas for features and labels not mentioned in the literature. Therefore, I want to interview experienced people from research and industry who use OSS libraries in their daily work. By interviewing a similar number of people from research and industry, I attempt to reduce bias in the resulting features and labels. To support the respondents, I plan to systematically prepare and present all features and labels found during the literature review. During the interviews, I also aim to interrogate the latest mechanisms, methods, and level of automation in monitoring maintenance activities of OSS libraries that are being applied during software development in research and industry.

The last step in answering RQ1 is to create a multi-class labeled dataset. Therefore, I plan to focus on frequently used package managers which store the code of their libraries publicly (e.g., GitHub, GitLab), such as PyPI, Maven and NPM \cite{xu2020reinventing}. For each package manager, I want to select libraries with low, medium and high popularity to reduce the bias in my dataset compared to previous approaches which mainly focused on highly popular libraries \cite{coelho2020github,li2022ossara,pashchenko2020vuln4real}. An extended dataset size might also positively influence the classification and prediction mentioned in section \ref{researchplanrq2} and \ref{researchplanrq3}.
All direct and transitive dependencies of the selected libraries should also be included in the dataset to be able to calculate the transitive impact in a later stage. For each selected library, I plan to extract the identified maintenance activity features and assign a label following a proposed labeling strategy. To better understand the resulting dataset, a descriptive analysis will be performed using unsupervised statistical methods such as multi-class clustering or principal component analysis (PCA).

\subsection{RQ2: Classifying Maintenance Activities}
\label{researchplanrq2}

Based on the labeled dataset from the previous RQ, I plan to extend and improve the state-of-the-art approaches for classifying OSS libraries in terms of their maintenance activity labels by considering direct and transitive dependencies. Therefore, I intend to conduct different research activities. First, I plan to carry out a feature important analysis and extend existing approaches to move from a binary to a multi-class classification approach using a random forest classifier, similar to \cite{coelho2020github}. This allows my results to be compared to previous work when only project-level features are considered.

Second, I plan to extend the state-of-the-art classification approaches by integrating the impact of direct and transitive dependencies in terms of their maintenance activities. Therefore, I want to follow the idea of \citeauthor{mujahid2021toward} to apply Google's page rank algorithm, which is suitable to capture the transitive impact within a network of dependencies \cite{rogers2002google}. In my case the page rank algorithm should capture aggregated maintenance activity features for each dependency within a network of OSS dependencies when classifying OSS libraries in terms of their maintenance activities. Also, different impact measurements could be evaluated to capture transitive effects, such as degree centrality, in-degree centrality, out-degree centrality and eigenvector centrality \cite{banks2022measuring}.

\subsection{RQ3: Predicting Maintenance Activities}
\label{researchplanrq3}

I plan to extend and improve the state-of-the-art approaches to predict the maintenance activities of OSS libraries considering direct and transitive dependencies. There are two important aspects when predicting maintenance activities. First, I want to test different ML methods (e.g., LSTM RNNs) which successfully proved to capture time series data. Second, I want to evaluate, how different time periods affect the prediction performance and confidence levels (e.g., 1, 3, 6, 9, 12 months), similar to \citeauthor{xia2022predicting} \cite{xia2022predicting}. 

Besides predicting project-level maintenance activity features of a library, the effect of direct and transitive dependencies should also be considered which is currently neglected by existing approaches. In contrast, I plan to predict a set of maintenance activity features for all libraries connected via direct and transitive links to a library under investigation. Afterwards, the classification can be performed as described in section \ref{researchplanrq2}, using the predicted features to classify the maintenance activity status in the future. 

\subsection{RQ4: Case Study on Effort Awareness}

I plan to evaluate the described methods in a case study focusing on effort awareness of application maintainers using a tool. The tool must be applicable in an industrial context providing two main functionalities: first, it should classify OSS dependencies into one of the many maintenance activity states identified in the mixed-method study. Second, it should predict the maintenance activities of OSS dependencies for different time periods. Both functionalities should take direct and transitive dependencies into account. 

I will try to find a collaboration partner from the industry who is willing to conduct a case study. Alternatively, if no industry partner is open to collaborate, the case study can be done using dependency data from OSS projects applied in the industry, such as Tensorflow, Jenkins, or Angular. Consequently, the study design remains independent of the data source.

The case study aims to evaluate the extent to which an application maintainer's effort to monitor the maintenance activities of OSS dependencies can be reduced using a tool. As baseline, I would calculate the number of libraries, including direct and transitive dependencies, that are theoretically required to be analyzed when manually monitoring the maintenance activities of all OSS dependencies. This baseline can be compared to the result of the tool-based approach, which describes a measure for effort savings. 

For example, a system containing 30 direct dependencies would lead to a baseline of 150 libraries when collecting all its direct and transitive dependencies. Imagine, out of these 150 libraries 20 are suspicious and should be analyzed more carefully by the application maintainer. In case my tool would report 15 out of 20 libraries as suspicious, the proposed tool would reduce the effort by 90\% compared to the manual approach, having an recall of 75\%. 

I argue, that recall is more important as evaluation metric compared to precision, because missing a positive case can lead to serious consequences. Contrary, one could oppose that precision might be more important to not overwhelm the user. In the end, it depends on the context and can be looked at from both sides.

To extend the evaluation of RQ4, more sophisticated effort awareness metrics known from other software engineering topics shall be applied, such as just-in-time defect prediction \cite{bludau2022feature}. In addition, effort awareness metrics could be translated into economic aspects to make the impact of my work more transparent for companies.

\subsection{Validity Threats and Mitigation Strategies}
Threats to validity are divided into the following four aspects \cite{runeson2009guidelines}: 

\textbf{Construct validity:}
To automatically classify OSS libraries in terms of their maintenance activities, I will construct a dataset containing maintenance activity data of OSS libraries from different package managers. To mitigate a potential threat to construct validity during dataset creation when selecting relevant features, I plan to define an appropriate termination condition to add additional features following \citeauthor{nickerson2013method} \cite{nickerson2013method}.

\textbf{Internal validity:} 
Due to the mixed-method approach, I am confident that most relevant features will be found that show relations to the maintenance activity labels identified. However, it is possible that aspects outside the scope have hidden relationships to maintenance activities that no one has ever thought of. 

\textbf{External validity:} 
As I will only analyze libraries from selected package managers, the findings might not directly be generalizable to other programming languages. However, the applied methods are generalizable to package managers of other programming languages if the identified features are available in these ecosystems.

Even though my focus is on industrial applications, the methods generalize to other contexts where OSS libraries are used during software development (e.g., research).

\textbf{Reliability:} 
To reduce the threat to reliability, all necessary artifacts to reproduce the results will be published and required data will be collected from publicly accessible sources (e.g., GitHub).

\section{Current Status}
Currently, I am working on the first research question (RQ1) which includes a mixed-method study to collect features and labels characterizing the maintenance activities of OSS libraries. Therefore, I am collecting state-of-the-art features and labels in a quantitative study using a literature review. Afterwards, new ideas for features and labels should be generated in a qualitative study using interviews. During these interviews, I also aim to interrogate the latest mechanisms, methods, and level of automation in monitoring maintenance activities of OSS dependencies that are being applied during software development in research and industry.

After a final set of features and labels is identified, I plan to define relevant libraries of popular package managers (e.g., NPM, PyPI, Maven) that will be used to collect maintenance activity data for the identified features. The results of RQ1, including features, labels, a descriptive analysis of the dataset, the dataset itself and the results of the interviews shall be published in a separate publication.
\section{Conclusion}

This paper intends to summarize the vision of my PhD project to identify suspicious OSS dependencies in applications in terms of their maintenance activities. I am confident that both research and industry can benefit from my project, as the adoption of OSS libraries is widespread nowadays, but too little attention is paid to potential security threats in this regard. My work can contribute to reduce the effort of application maintainers since only a small portion of the whole library set must be analyzed. In addition, my work supports application security by continuously having well-maintained libraries in a system preventing the situation of a library being affected by a vulnerability without active community support. To create the greatest possible value for other researchers, I plan to make all artifacts from future publications accessible.

\bibliographystyle{ACM-Reference-Format}
\bibliography{bibliography}

\appendix

\end{document}